\documentclass[10pt]{article}
\pdfoutput=1
\usepackage{jcappub}

\usepackage{url} %
\usepackage{physics} 

\usepackage{multirow}
\usepackage{booktabs}
\usepackage{placeins}

\newcommand{\unit}[1]{\,\mathrm{#1}}
\newcommand{\factor}[3]{\left(\frac{#1}{#2\unit{#3}}\right)}
\renewcommand{\vec}[1]{\boldsymbol{#1}}

\def\epsilon{\varepsilon}
\def\theta{\vartheta}

\newcommand{\apj}{{Astrophys.\ J. }}

\newcommand{\araa}{{Ann.\ Rev.\  Astron.\ Astrophys. }}

\newcommand{\mnras}{{Mon.\ Not.\ Roy.\ Astron.\ Soc. }}

\title{Reacceleration of charged dark matter}

\author{M. Kachelrie{\ss} and}
\author{J. Tjemsland}

\affiliation{Institutt for fysikk, NTNU, Trondheim, Norway
  \vskip 0.3cm
  Received: June 15, 2020\\
  Accepted: July 31, 2020
}


\abstract{
Charged particles scattering on moving inhomogenities of the magnetised
interstellar medium can gain energy through the
process of second-order Fermi acceleration. This energy gain depletes in
turn the magnetic wave spectrum around the resonance wave-vector $k\sim 1/R_L$,
where $R_L$ is the Larmor radius of the charged particle. This energy transfer
can prohibit the cascading of magnetic turbulence to smaller scales, leading
to a drop in the diffusion coefficient and allowing the efficient exchange of
charged dark matter particles in the disk and the halo. As a result,
terrestial limits from direct detection experiments apply to charged dark
matter. Together with the no-observation of a drop in the diffusion
coefficient, this excludes charged dark matter for
$10^3\unit{GeV}\lesssim m/q \lesssim 10^{11}\unit{GeV}$, even if the
charged dark matter abundance is only a small part of the total relic
abundance.
}

\begin{document}

\maketitle

\section{Introduction} 
\label{sec:introduction}

There is overwhelming evidence for the presence of a non-baryonic component
in the matter budget of the Universe through its gravitational effects. If
this component consists of particles having only gravitational interactions,
there would be little hope for their detection. Therefore it is commonly assumed
that they participate in part of the gauge interactions of the standard model
(SM). In order to keep the particles dark and collisionless, typically the
weak interaction is chosen as coupling to the SM particles. However, no new
weakly interacting particles have been detected yet, despite of intense
searches. As a consequence, the interest in alternatives has been growing.
One option is the case of millicharged dark matter (DM), where the
DM particle has a small, but non-zero electric charge $qe\ll e$,
thereby providing a coupling via the photon to the SM
particles~\cite{Ignatiev:1978xj}. The smallness
of $q$ could be explained, e.g., by the kinetic mixing of the dark photon
of an extra U(1) symmetry with the SM
photon~\cite{Okun:1983vw,Georgi:1983sy,Holdom:1985ag,Dobroliubov:1989mr,Fayet:1990wx}. Alternatively, DM with charges of order one could form heavy
bound states~\cite{DeRujula:1989fe}.

The case of millicharged DM has obtained increased attention lately following
the reported discrepancy between predictions for the 21\,cm absorption line
induced by early stars and observations by the EDGES
collaboration~\cite{Bowman:2018yin}. It has been suggested that this anomaly
is caused by the cooling of baryons at redshift $z\simeq 17$ in baryon-dark
matter interactions with a massless mediator~\cite{Barkana:2018lgd}. Dark
matter with a tiny electric charge $q$ would be a natural candidate for such a
new cooling agent. Unfortunately, a scenario where the EDGES observations are
explained by millicharged DM that explains all the observed DM abundance is
inconsistent with bounds from the cosmic microwave
background~\cite{Kovetz:2018zan}. Therefore, DM theories in which there is a
sub-dominant charged component, such as theories including a completely
hidden sector have attracted attention, see e.g.\
Refs.~\cite{Munoz:2018pzp,Hertzberg:2019bvt,Liu:2019knx,Kahlhoefer:2020rmg}
for recent examples and Refs.~\cite{Fabbrichesi:2020wbt,Filippi:2020kii}
for reviews. 

Cosmological observations and direct detection experiments put strong 
constraints on the possible mass-to-charge ratios of
DM~\cite{Davidson:1991si,McDermott:2010pa}. However, Chuzhoy
and Kolb \cite{Chuzhoy:2008zy} claimed in 2009 that the direct detection
constraints on charged DM are invalid in a wide range of $m/q$ values:
They argued that the regular\footnote{Note that
  Liouville's theorem implies that the intensity of charged DM is constant
  along a trajectory in the magnetic field. Thus a shielding effect analogue
  to the geomagnetic cutoff would require, e.g., a dipole component of the
  GMF.} Galactic magnetic field (GMF) in the disk prevents charged DM
particles with $m/q\lesssim 10^{11}\unit{GeV}$ from entering the plane from the
halo. Taking in turn into account the acceleration of charged DM particles in
the disk by shock fronts of supernova remnants and the subsequent loss
of energetic DM particles into the halo, they argued that charged DM with
$10^5q^2\lesssim m/\mathrm{GeV}\lesssim 10^{11} q $ is expelled from the
disk and evades thereby the bounds
from terrestrial direct DM searches. This range was later reduced to 
$m/q\lesssim 10^9\unit{GeV}$ in Ref.~\cite{SanchezSalcedo:2010ev} by taking 
into account the effect of the turbulent magnetic field and the
non-homogeneity of the background field.
The authors of Ref.~\cite{Li:2020wyl} argued that the injection of charged DM
into diffusive shock acceleration is suppressed, implying that these particles
are not effectively accelerated in shock fronts of supernova remnants.
The most recent and complete study treating charged DM as
diffusive cosmic rays has been preformed in Ref.~\cite{Dunsky:2018mqs}.
In this analysis, it was found that there is a substantial amount of charged
DM present in the Galactic disk today, some of which have recently been
accelerated. This was used to set strong  constraints on the possible
$(m, q)$ parameters.

In this work, we consider a generic DM particle with mass $m$ and charge $qe$.
As any electrically charged particle, charged DM  will scatter on the
inhomogeneities of the turbulent GMF. These inhomogeneities are moving with
typical velocities of tens of km/s with respect to the Galaxy, either
because static turbulence is advected alongside the plasma or because
the turbulence consists of travelling Alfv\'en waves. Both cases lead to
second-order Fermi acceleration of charged particles including charged DM.
The energy gained by the charged DM depletes in turn the magnetic wave spectrum
around the resonance wave-vector $k_{\rm res}\sim 1/R_L$, where $R_L$ is the
Larmor radius of the charged DM particle.
If this energy drain is larger than the power injected into the turbulent
GMF at the injection wave-number $k_{\min}$, the cascading of magnetic
turbulence from $k_{\min}$ to larger wave-numbers is stopped at $k_{\rm res}$.
In this case, all fluctuations in the inertial range above
$k_{\rm res}$ are  missing. As a result, charged DM from the halo can enter
the disk and the limits from direct DM searches can be applied.
Together with the no-observation of a drop in the diffusion
coefficient, this excludes charged DM for a wide range of masses
and charges.

\section{Alfv\'en waves and diffusion} 
\label{sec:diffusion_equation}

\subsection{Diffusion equations}

The propagation of charged particles through the magnetized interstellar medium
(ISM) filling the Milky Way can be described phenomenologically as a
combination of diffusion and advection using a Fokker-Planck
equation~\cite{1969ocr..book.....G,Skilling:1975a}: The scattering of charged
particles on the inhomogeneities of the turbulent GMF
leads to diffusion terms in the evolution equation for the
phase space density $f(\vec x,\vec p)$ of the charged DM particle,
\begin{equation}
    \pdv{f}{t} = Q
    + \vec\nabla(D(p)\vec\nabla f) 
    + \frac{1}{4\pi p^2}\pdv{}{p}~\left(4\pi p^2 D_{pp}(p) \pdv{f}{p}\right) 
    +\ldots.
    \label{eq:diff_f}
\end{equation}
Here, $D$ and $D_{pp}$ parametrise diffusion in position and momentum space,
respectively, while $Q$ is a source term and $p=|\vec p|$ the momentum of the
charged DM particle. Moreover, we assumed for simplicity that the diffusion is
isotropic, i.e.\ we replaced the diffusion tensor $D_{ij}(\vec p)$ by the scalar
diffusion coefficient $D(p)$. Additionally, charged particles are advected
with the plasma.

Alfv\'en waves are solutions of the MHD equations which propagate approximately
parallel to the magnetic field lines with the Alfv\'en velocity%
\footnote{We use Gaussian units and set $c=1$.}
$v_A=B_0/\sqrt{4\pi\rho}\simeq 30$\,km/s, where $\rho$ is the density of
the plasma and $B_0$ the strength of the magnetic background field.
Charged particles are dynamically coupled to the ISM via
the self-generation of, and the scattering on, Alfv\'en waves. 
Thus, the usual ``test-particle''
approach, where charged particles propagate in a prescribed static background
(or given 
diffusion coefficients $D$ and $D_{pp}$ in the diffusion picture) is
in general not valid. Instead, one must check if their back-reaction on
the turbulent magnetic field at scales comparable to their Larmor radius
is negligible.

The rate at which energy is transferred between Alfv\'en waves and
charged particles is given by\footnote{We assume that the forward and backward
 scattering rates are the same.}
\begin{equation}
  \Gamma_\mathrm{growth} = \frac{16\pi^2}{3}\frac{p^4v_A}{kW_k(t)B_0^2}
  \left(v\hat{\vec n}\cdot\vec\nabla f - 
  \frac{\pi}{2}m\Omega_0v_AkW_k(t)\pdv{f}{p}\right),
  \label{eq:growth}
\end{equation}
where $\Omega=qeB_0/m$ is the cyclotron frequency of a particle with mass
$m$ and charge $qe$,  $v$ its velocity, and $W_k$ denotes the spectral density
of turbulent field modes with wave-vector $\vec k$~\cite{Skilling:1975c}.
This interaction proceeds resonantly, such that $k$ equals
$k_{\rm res}\simeq\Omega/v\mu$ with $\mu=\cos\theta$ as the cosine of
the pitch angle~\cite{Melrose:1968}.
We normalise the spectral density of turbulent field modes $W_k$ such that
$\eta_B\equiv\delta B^2/B_0^2=\int\dd{k}W(k)$, where $\eta_B$ denotes
the ratio of the energy density in the turbulent and the regular magnetic
field. Sources of this turbulence are the mechanical ``stirring''
of the plasma and the generation of Alfv\'en waves by charged particles.
In the former case, stellar winds, supernova shocks and the differential
rotation of the Milky Way inject energy into the ISM on scales of tens
of parsecs, which then cascades down to smaller scales through the formation
of smaller and smaller eddies. This energy cascade in the inertial range
can be  modelled as a diffusion process in $k$ space~\cite{1979ApJ...230..373E},
\begin{equation}
  \pdv{W_k(t)}{t} = \pdv{}{k}~\left(D_{kk}\pdv{W_k(t)}{k}\right) - 
  \Gamma_\mathrm{growth}(k, t)W_k(t) + q_W,
  \label{eq:diff_W}
\end{equation}
with the diffusion coefficient~\cite{doi:10.1029/JA095iA09p14881}
\begin{equation}
  D_{kk}=C_Kv_Ak^{\alpha_1}W_k(t)^{\alpha_2}
\end{equation}
and $C_K\simeq 0.052$~\cite{refId0}. The injection occurs via the source
term $q_W\propto \delta(k-2\pi/L_{\max})$ at the outer scale $L_{\max}\simeq 100\unit{pc}$.
The parameters $\alpha_i$ are chosen as  $\alpha_1=7/2$ and $\alpha_2=1/2$
such that the power-law spectrum
\begin{equation}
  W(k)=W_0\left(\frac{k}{k_{\min}}\right)^{-s},\qquad s = 
  \frac{\alpha_1-1}{\alpha_2 + 1},\qquad W_0=(s-1)L_{\max}\eta_B,
  \label{eq:wave_power_kolmogorov}
\end{equation}
obtained as steady-state solution for $\Gamma_\mathrm{ex}=0$
is a Kolmogorov spectrum. The wave-number $k_{\min}=2\pi/L_{\max}$
is determined by the injection scale $L_{\max}$ of the turbulence.

Equations~\eqref{eq:diff_f} and \eqref{eq:diff_W} form a set of coupled
differential equations that must be solved iteratively in order to compute
the time evolution of both the phase-space density $f$ of charged DM and the
spectrum $W_k$ of magnetic field fluctuations.

\subsection{Resonance condition}

Since the interaction~(\ref{eq:growth}) between charged particles and
Alfv\'en waves proceeds resonantly, we have to check that the resonant
wave-vector $k_{\rm res}$ of the charged DM particle is contained in the inertial
range $[k_{\min}:k_{\max}]$ of the turbulent cascade. Here,
$k_{\max}=2\pi/L_{\min}$ is given by the dissipation scale where the
turbulent energy is converted into heat. While observations show that
fluctuations extend down at least to $10^9$\,cm~\cite{Armstrong:1995zc,Chepurnov:2009ym},
it is theoretically expected that the dissipation scale corresponds to
the proton or even electron Larmor
radius~\cite{Boulanger:2018zrk,Schekochihin:2007mw}. For concreteness,
we will use in the following the proton Larmor radius as the dissipation
scale.

The resonant wave-number for the momentum $p=\gamma m v$ is~\cite{Melrose:1968}
\begin{equation}
    k_\mathrm{res}=\frac{\Omega m}{p}\frac{1}{\mu \pm v_A/v} \geq 
    \frac{\Omega m}{p}\frac{1}{1 \pm v_A/v}.
\end{equation}
We first check that the condition $k_{\rm res}<k_{\max}$ is satisfied.
Using that the maximum wave-number is limited by the proton cyclotron
frequency, $k_\mathrm{max} \simeq \Omega_{\rm p}/v_A$, the momentum that
can resonate with Alfv\'en waves must satisfy
\begin{equation}
    p \gtrsim mv_A\left(\frac{qm_{\rm p}}{m} \mp 1\right)
\end{equation}
in the non-relativistic case, $p=\gamma mv\approx mv$. In the for us
interesting limit $m/q\gg m_{\rm p}$, this reduces\footnote{The choice of
  sign means that the charged DM can resonate only with one of the two
  polarisation states of the Alfv\'en waves.} to
$p \gtrsim mv_A$. Since the virial velocity of the
DM, $v_\mathrm{vir}\sim 300\unit{km/s}$, is much larger than the
Alfv\'en velocity, the majority of the charged DM particles fulfills this
condition.

Next we examine when the condition $k_{\rm res}>k_{\min}$ is satisfied.
Setting for simplicity $\mu=1$, it is $k_{\rm res}=\Omega/v=1/R_L$
with $R_L = {\gamma mv}/{qeB_0}$ as the Larmor radius.  Thus, requiring 
\begin{equation}
  R_L \simeq 1.08\times 10^{-3}\unit{pc}\factor{v_{\rm vir}}{300}{km/s}
     \factor{B_0}{1}{\mu G}^{-1}\factor{m/q}{10^6}{GeV} \lesssim L_{\max}
\end{equation} 
with $L_{\max}=100$\,pc gives the constraint
\begin{equation}
    m/q \lesssim 10^{11}\unit{GeV}.
\end{equation}

\subsection{Diffusion coefficients}
\label{sec:diffusion_coefficient}

The diffusion coefficients can be written as
\begin{equation}  \label{diff0}
  D = \frac{1}{3}\frac{v^2}{\nu_++\nu_-}\qquad\mathrm{and}\qquad
  D_{pp} = \frac{4}{3}\gamma^2m^2v^2v_A^2\frac{\nu_+\nu_-}{\nu_+ + \nu_-},
\end{equation}
where $\nu_\pm=\Omega kW(k)/\gamma$ are the collision frequencies of the 
forward and backward propagating Alfv\'en waves at the resonance $k_{\rm res}$,
and $\gamma$ is the Lorentz facor of the charged
particle~\cite{Skilling:1975c}. 
Assuming again that the forward and backwards rates are equal and inserting
the expression~\eqref{eq:wave_power_kolmogorov} for the wave spectrum, the
diffusion coefficients can be rewritten as
\begin{equation}
  D = \frac{vL_{\max}}{3}[\eta_B(s-1)]^{-1}
  \left(\frac{R_L}{L_{\max}}\right)^{2-s}
  \qquad\mathrm{and}\qquad 
  D_{pp} = \frac{1}{3}\gamma^2m^2v_A^2
  L_{\max}\eta_B(s-1)
  \left(\frac{R_L}{L_{\max}}\right)^{s-2},
\end{equation}
where
$s=(\alpha_1-1)/(\alpha_2 + 1) = 5/3$. Choosing then
$\eta_BL_{\max}\simeq 1 \unit{pc}$, this
framework is consistent with the commonly used parametrisation
$D = D_0v(p/m)^\delta$ with $\delta = 2-s$ and
\begin{equation}
    D_0 = 3\times 10^{28}\unit{\frac{cm^2}{s}}\left[
    \factor{m/q}{10^6}{GeV}\factor{B}{1}{\mu G}^{-1}\right]^{2-s} .
\end{equation}

\section{Power taken by millicharged dark matter} 
\label{sec:power_taken_by_millicharged_dark_matter}

The power density required for the reacceleration of cosmic rays is discussed
in Ref.~\cite{Thornbury:2014bwa}. There, it is found that $\simeq 10\%$ of
the power in the cosmic ray proton spectrum comes from reacceleration
and affects mainly the non- and mildly relativistic part of the cosmic
ray spectrum. The same process must occur for charged DM, which means that
its effect is potentially important.
Following the procedure in Ref.~\cite{Thornbury:2014bwa} and using the
diffusion coefficients given in section~\ref{sec:diffusion_coefficient},
we obtain as the reacceleration power density 
\begin{equation}
    P_R \approx \frac{1}{9}(4-\delta)\frac{v_A^2}{D_0} m
    \int_0^\infty \dd{p} 4\pi p^2\left(\frac{p}{m}\right)^{1-\delta} f(p),
\label{eq:power_equation}
\end{equation}
where $f(p)$ is the momentum distribution of the particles. For comparison,
we will consider protons with $f(p)\propto (p/m)^{-\gamma}$,
\begin{equation}
    P_R^\mathrm{protons} \simeq 
    0.56\unit{eV/cm^3}  \;    \frac{v_A^2}{{D_{0, \mathrm{proton}}}}.
\end{equation}
We consider a standard Maxwellian phase space density for the momentum
distribution of the charged DM particles,
\begin{equation}
  f(p, z) = (2\pi \epsilon^2)^{-3/2}
  \exp{-\frac{p^2}{2\epsilon^2}}n(z),
  \label{eq:maxwellian}
\end{equation}
with $\epsilon = \gamma m v_\mathrm{vir}/\sqrt{2}$ and
$n(z)= n_0\simeq 0.3\unit{(GeV/}m)/\mathrm{cm^3}$ as the local charged DM
density. Thus, we obtain
\begin{equation}
  \frac{P_R}{P_R^\mathrm{protons}} =4\times 10^6
  \factor{m}{10^6}{GeV}^{2/3}q^{1/3}
  \factor{v\gamma}{300}{km/s}^{1/2}
  \factor{B}{1}{\mu G}^{1/3}.
\end{equation}
This means that the power density going into the reacceleration of charged
DM is potentially large, exceeding formally the available power density
injected into the turbulent ISM.

\section{Key rates} 
\label{sec:key_rates}

\begin{figure}[htbp]
  \centering
  \includegraphics[scale=1]{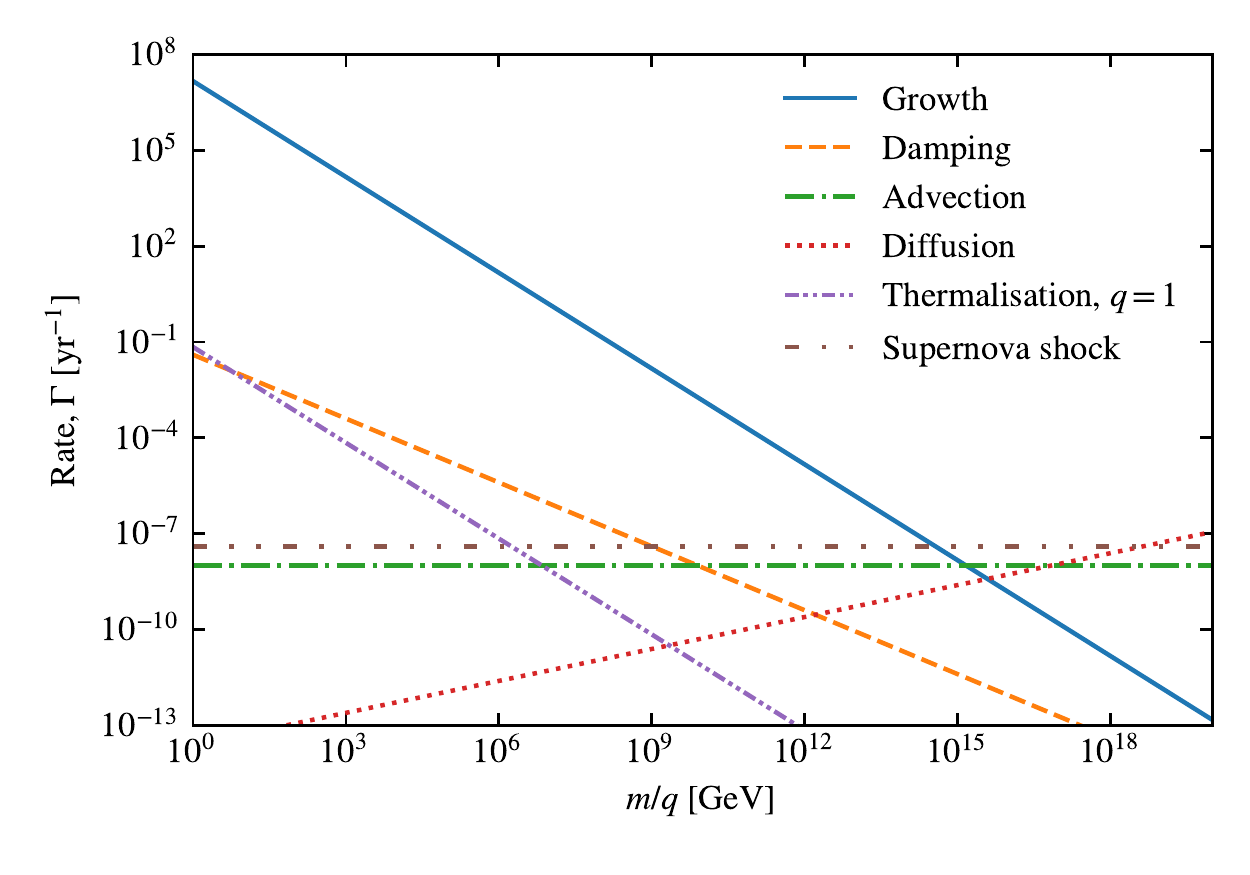}
  \caption{Various rates as a function of the charge-to-mass
  ratio of the charged DM.}
  \label{fig:rates}
\end{figure}

From the estimate in the previous subsection, it is clear that the
reacceleration of charged DM has the potential of  seriously disturbing
the ISM. In order to understand the potential consequences, we have
to estimate the relevant time-scales of the problem, which will be
discussed in this section. We will focus on the case where the entire
relic density consists of charged DM, but we will comment on the case of a
subdominant component in the next section. A summary of the results is shown
in Fig.~\ref{fig:rates}.

\subsection{Growth rate}

In order to estimate the growth rate given in Eq.~\eqref{eq:growth}, we
consider a Maxwellian phase space distribution \eqref{eq:maxwellian}.
That is, the momentum distribution becomes $4\pi p^2 f(p, z)$ so that the
number of particles in the momentum range $(p, p+\dd{p})$ is
$4\pi p^2 f(p, z)$.

With $\partial_z f\sim [n_0/H]\cdot[f/n(z)]$ and choosing $H\sim 3\unit{kpc}$
as the half-height of the confinement region, the growth rate induced by the 
spatial gradient is
\begin{equation}  \label{1a}
\begin{aligned}
  \Gamma_\mathrm{growth}^z &= \frac{2.4}{\unit{yr}} 
    \factor{m/q}{10^6}{GeV}^{-2/3}
    \factor{v}{300}{km/s}^{4/3}\gamma^{1/3}
    \exp{-\frac{p^2}{p_\mathrm{vir}^2}}\\
    & \factor{B_0}{1}{\mu G}^{-4/3}
    \factor{v_A}{30}{km/s}
    \factor{\eta_B}{0.05}{}^{-1}
    \factor{L_{\max}}{50}{pc}^{2/3}
    \factor{H}{3}{kpc}^{-1} .
\end{aligned}
\end{equation}
Likewise, using $n(z)\sim n_0$, the momentum gradient leads to the growth rate
\begin{equation}  \label{1b}
  \Gamma_\mathrm{growth}^p = \frac{40}{\unit{yr}}
    \factor{m/q}{10^6}{GeV}^{-1}
    \frac{p}{p_\mathrm{vir}}
    \exp{-\frac{p^2}{p_\mathrm{vir}^2}}
    \factor{B_0}{1}{\mu G}^{-4/3}
    \factor{v_A}{30}{km/s}^2.
\end{equation}
Since we in our picture start without any spatial gradient and 
the corresponding growth rate $\Gamma_\mathrm{growth}^z$ is small
even assuming a large gradient,
it can be neglected relative to $\Gamma_\mathrm{growth}^p$.
It is apparent from Fig.~\ref{fig:rates} that the growth rate is
dominant for $m/q\lesssim 10^{15}\unit{GeV}$,
where it has been evaluated using $p=p_{\rm vir}$ and the numerical
values of the astrophysical parameters given in Eqs.~(\ref{1a})
and~(\ref{1b}).

\subsection{Damping, diffusion and advection rates}

The relevant rates in the diffusion equations describing the propagation of
charged DM particles and their interaction with the wave spectrum are the
damping rate $\Gamma_\mathrm{damp}\sim D_{kk}/k^2$, the advection rate
$\Gamma_\mathrm{adv}\sim v_{\rm adv}/H$ and the diffusion rate
$\Gamma_\mathrm{diff}\sim 2D/H^2$. Numerically, we can estimate these rates as
\begin{align}
  \Gamma_\mathrm{damp} &= 
    \frac{4\times 10^{-6}}{\unit{yr}} 
    \factor{m/q}{10^6}{GeV}^{-2/3}
    \factor{v}{300}{km/s}^{-2/3}\gamma^{-2/3}\notag\\&\qquad\qquad\qquad\times
    \factor{B_0}{1}{\mu G}^{2/3}
    \factor{L_{\max}}{50}{pc}^{-1/3}
    \factor{\eta_B}{0.05}{}^{1/2}
    \factor{v_A}{30}{km/s},\\
  \Gamma_\mathrm{diff} &=
    \frac{2\times 10^{-12}}{\unit{yr}}
    \factor{m/q}{10^6}{GeV}^{1/3}
    \factor{v}{300}{km/s}^{4/3}\gamma^{-2/3}
    \factor{B_0}{1}{\mu G}^{-1/3}
    \factor{H}{3}{kpc}^{-2},\\
  \Gamma_\mathrm{adv} &=
    \frac{1\times 10^{-8}}{\unit{yr}}
    \factor{v_{\rm adv}}{30}{km/s}
    \factor{H}{3}{kpc}^{-2}.
\end{align}
Since charged DM is non-relativistic, advection dominates over diffusion
except for the largest $m/q$ values considered in Fig.~\ref{fig:rates}.
However, all three rates are well below the growth rate in the range
where the resonance condition is satisfied.

\subsection{Thermalisation rate}

The thermalisation time scale for a  particle with charge
$q_1e$, mass $m_1$ and velocity $v$ passing through a medium consisting of 
particles with charge $q_2e$, mass $m_2$ and density $n$ is given by
\begin{equation}
  t_c = \frac{m_1m_2v^3}{8\pi q_1^2q_2^2e^4n_e\ln \Lambda} ,
\end{equation}
where we use $\ln\Lambda\sim 20$ as Gaunt factor~\cite{2011hea..book.....L}. The
thermalisation is dominated by the warm ionized medium which has fractional
volume $f_\mathrm{WIM}=0.15$, electron density $n_e=0.2\unit{cm}^{-3}$ and
temperature $8\times 10^3\unit{K}$~\cite{2005ARA&A..43..337C}. Since the
velocity of the
charged DM, $v_\mathrm{vir}\sim 300 \unit{km/s}$, is smaller than the electron 
velocity in the warm ionized medium, $v_e\sim 600\unit{km/s}$, the charged DM will
thermalise at a rate
\begin{equation}
  \Gamma_\mathrm{therm} = \frac{1}{t_c}\sim\frac{7\times 10^{-8}}{\unit{yr}}
  \factor{m/q^2}{10^6}{GeV}^{-1}\factor{v_e}{600}{\unit{km/s}}^{-3}
  \factor{n_e}{0.2}{cm^3}\factor{f_\mathrm{WIM}}{0.15}{}
\end{equation}
in the thin disk. As visible from Fig.~\ref{fig:rates}, this rate is much
smaller than the other relevant rates. In addition, thermalisation only occurs
in the thin Galactic disk, but the charged DM spend most of its time outside this
region. Thus, thermalisation can be neglected in the present work.

\subsection{Supernova shock encounter rate}

The effect of supernova remnants on charged DM was studied in
Ref.~\cite{Dunsky:2018mqs}. The expected rate at which charged DM
particles in the Galactic disc will encounter supernova shocks
is~\cite{Dunsky:2018mqs}
\begin{equation}
  \Gamma_\mathrm{SH} = \frac{4\times 10^{-8}}{\unit{yr}}
  \factor{R_\mathrm{max}}{40}{pc}^3
  \factor{R_\mathrm{disc}}{15}{kpc}^{-2}
  \factor{H_\mathrm{disc}}{300}{pc}^{-1}
  \factor{\Gamma_\mathrm{SN}}{0.03}{yr^{-1}},
\end{equation}
This rate is much smaller than the growth rate for $m/q\lesssim 10^{15}\unit{GeV}$. 
Thus, the acceleration of charged DM by supernova shocks can be
neglected.

\subsection{Injection rate of turbulence}

The turbulence is injected at scales $L_{\max}\sim 50\text{--}100\unit{pc}$
through the source term $q_W\propto \delta(k-2\pi/L_{\max})$. 
Without the presence of charged DM, a Kolmogorov spectrum will develop. Thus,
the source term can be found as
\begin{equation}
  q_W(k) = \frac{5\times 10^{-11}}{\unit{yr}} 
  \factor{\eta_B}{0.05}{}^{3/2}
  \factor{v_A}{30}{km/s}\factor{L_{\max}}{50}{pc}^{-1}\delta(k-2\pi/L_{\max}).
  \label{eq:injection}
\end{equation}
Meanwhile, the rate for the absorption of wave power by charged DM can
be estimated as
\begin{equation}
\begin{aligned}
  \int_{k_{\rm min}}^{k_{\rm max}}&\dd{k} \Gamma_\mathrm{growth}W \simeq 
  -\int_{\gamma mv_A}^\infty\dd{p}\frac{qeB_0}{p^2c} \Gamma_\mathrm{growth}W\\
  &=\frac{2\times 10^{-4}}{\unit{yr}} 
  \factor{m/q}{10^6}{GeV}^{-1}
  \factor{v_\mathrm{vir}}{300}{km/s}^{2/3}\gamma^{2/3}
  \factor{\eta_B}{0.05}{}
  \factor{L_{\rm max}}{50}{pc}^{-2/3}
  \factor{v_A}{30}{km/s}^{2}.
\end{aligned}
\label{eq:max_absorption}
\end{equation}
In the second step, we extended the integration region from the resonance
up to infinity, which is admissible since the gradients of $f$ are 
strongly peaked. Moreover, we used $mv_A/p_\mathrm{vir}\sim 0.1$ to obtain
a numerical value of the
resulting Gamma function. According to this simple estimate, the absorption
rate of turbulence is larger than the injection rate for
$m/q\lesssim 10^{12}\unit{GeV}$, i.e.\ in all the range
where the resonance condition is satisfied.

\section{Time evolution and observable consequences}
\subsection{Time evolution}
\label{sec:solution_of_the_diffusion_equations}

In order to better understand the process of cascading and absorption of
wave power, we solve the time-dependent diffusion equations~\eqref{eq:diff_f} 
and \eqref{eq:diff_W} using the Crank-Nicholson scheme. As initial condition,
we use the Maxwellian phase space distribution~\eqref{eq:maxwellian} with
$n(z)=n_0$. From the discussions in the previous section, we know that the
reacceleration is dominant, and we therefore neglect convective and spatial
diffusive terms\footnote{
  These terms can in principle be taken into account by introducing a
  ``leaky-box'' loss term $f/T$ and source term
  $f_\mathrm{vir}/T$.} in the diffusion equation~\eqref{eq:diff_f}.
Likewise, the growth due to the spatial gradient in Eq.~\eqref{eq:growth} is
subdominant and can be neglected.

\begin{figure}[htbp]
  \centering
  \includegraphics[scale=1]{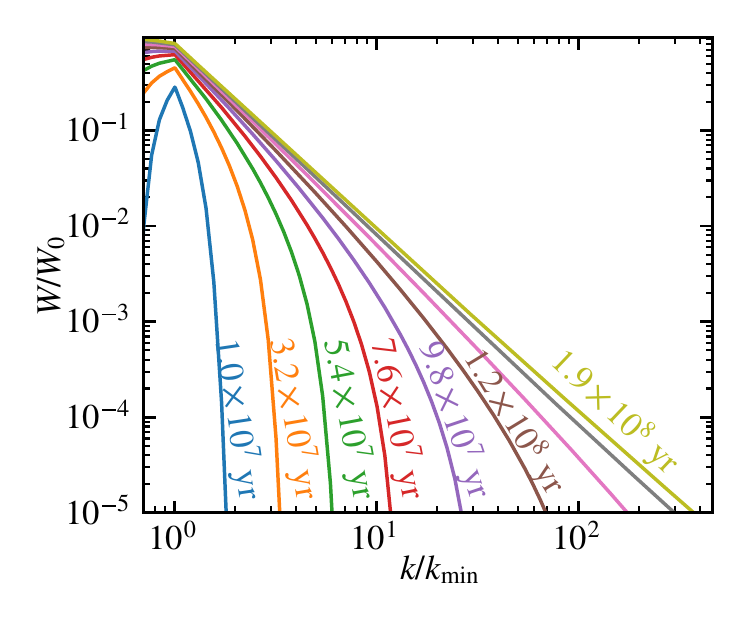}
  \includegraphics[scale=1]{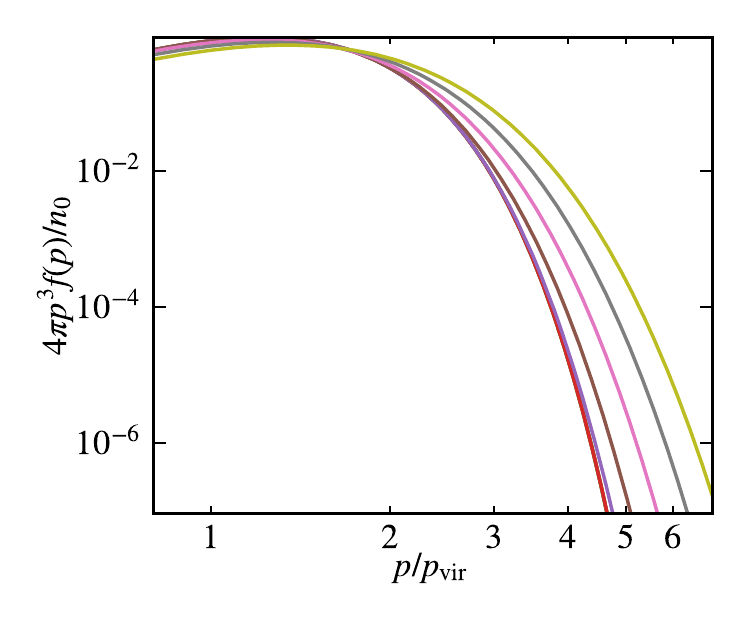}

  \vspace{-.45cm}
  \includegraphics[scale=1]{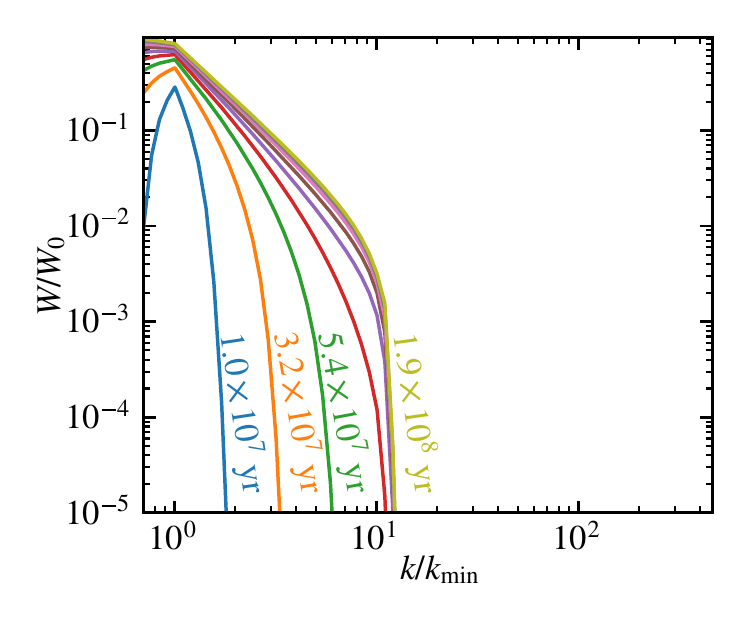}
  \includegraphics[scale=1]{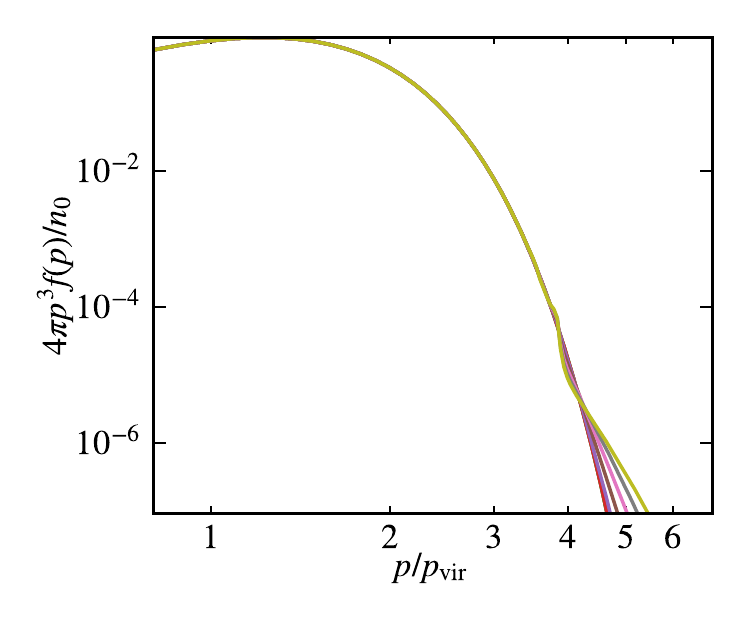}

  \vspace{-.45cm}
  \includegraphics[scale=1]{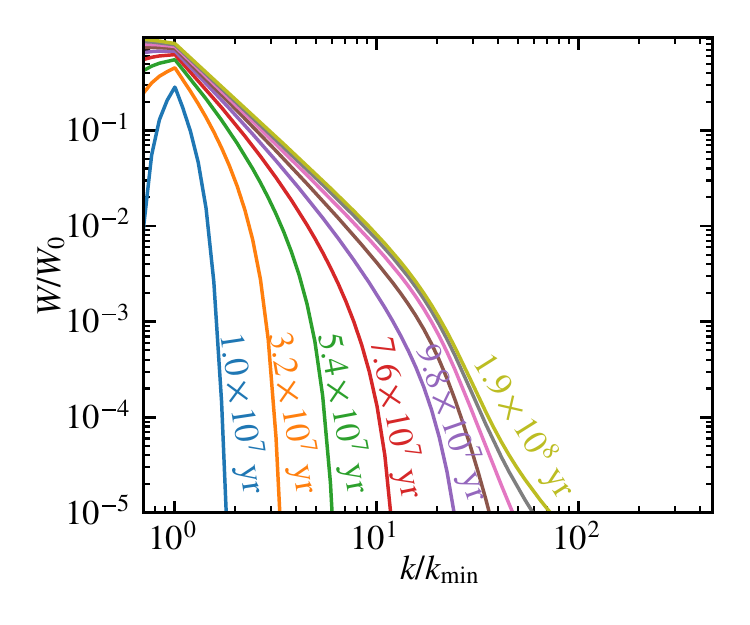}
  \includegraphics[scale=1]{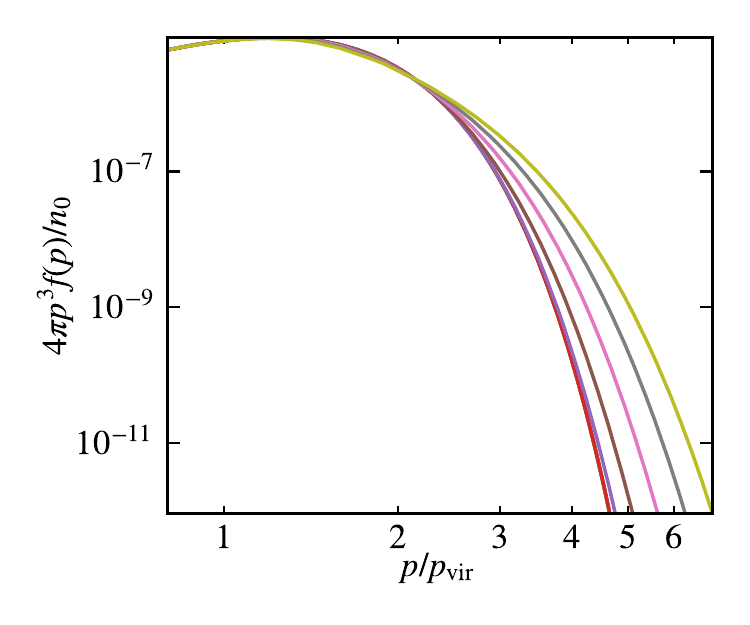}
  \caption{Time evolution of the wave power (left column) and the phase space
  density (right column).  
  In the first row the growth rate is set to zero, while it is
  included in the second row. In the third row, the growth rate is included 
  and $f_{\rm cDM}=10^{-5}$.}
  \label{fig:diffusion_results}
\end{figure}

For concreteness, we consider $m=10^6\unit{GeV}$ and $q=10^{-3}$ as an example.
Moreover, we assume first that the charged DM abundance $\Omega_{\rm cDM}$ equals
the total DM abundance $\Omega_{\rm DM}$, i.e.\
$f_{\rm cDM}=\Omega_{\rm cDM}/\Omega_{\rm DM}=1$.
In the first row of figure~\ref{fig:diffusion_results}, we show how the wave
power cascades to larger wave-numbers and how this affects the phase space
density of the charged DM setting by hand the growth term
$\Gamma_\mathrm{growth}$ in Eq.~(\ref{eq:growth}) to zero: As the time increases,
the turbulence injected around the scale $k_{\min}$ cascades to larger
and larger wave-numbers, forming a Kolmogorov power-law spectrum,
while the phase-space density of charged DM develops an increasing
high-energy tail. Switching on the growth term, however, the turbulent
cascade stops at  $k\sim qeB_0/(p_\mathrm{vir}\times\mathrm{few})$,
as shown in the second row. Note that in this case the momentum distribution
of charged DM does not change drastically. 

For a fully developed Kolmogorov spectrum we know from
Eqs.~\eqref{eq:injection} and \eqref{eq:max_absorption} that all wave power is
absorbed by the charged DM for $f_{\rm cDM}^{-1}m/q\lesssim 10^{12}\unit{GeV}$. 
For $m=10^6\unit{GeV}$ and $q=10^{-3}$ this transition occurs at
$f_{\rm cDM}\sim 5\times 10^{-4}$. 
To visualise the effect of partial absorption, we show in the third row of
Fig.~\ref{fig:diffusion_results} the solution with $f_{\rm cDM}=10^{-5}$. The
wave-spectrum $W(k)$ has now a drop at the resonance momentum, but
recovers at larger $k$. This drop leads via Eq.~(\ref{diff0}) to a
corresponding jump in the diffusion coefficient $D(p)$. 

\subsection{Observable consequences} 
\label{sec:observable_consequences_of_reacceleration}

Turbulence injected at large scales $L_{\max}\sim 50\text{--}100\unit{pc}$
cascades to smaller wave lengths creating in the inertial range a power-law
spectrum. In the presence of charged DM, however, the wave power will be
absorbed when the cascade reaches $k_\mathrm{vir}\sim q eB_0/p_\mathrm{vir}$.
Since the growth rate is for $f_{\rm cDM}=1$ much larger than the damping rate,
the entire wave energy will be absorbed such that no waves can cascade above
$\sim k_\mathrm{vir}$. Thus, for charged DM with 
$m/q\lesssim 10^{11}\unit{GeV}$ the cascading will stop at $k_\mathrm{vir}$.
Therefore no charged particles with momenta above  $p_\mathrm{vir}$ 
be able to resonate with Alfv\'en waves. Effectively, this would lead to a
sudden drop in the diffusion coefficient which is not observed in the
cosmic ray spectra~\cite{Kachelriess:2019oqu}. Similarly, cosmic rays
with momenta above  $p_\mathrm{vir}$ would not be isotropised by the
GMF, in contradiction to the very low level of anisotropy
observed~\cite{Kachelriess:2019oqu}. Using the absence of anisotropies
and a sudden
drop in spectra observed above $\sim 0.1\unit{GeV}$, one can exclude 
$10^3\unit{GeV}\lesssim m/q \lesssim 10^{11}\unit{GeV}$.
Additionally, charged DM is subject to upper limits set by terrestrial direct
detection experiments, since in the absence of resonant Alfv\'en waves charged
DM particles are exchanged freely between the Galactic disk and halo.

\begin{figure}[htbp]
  \centering
  \includegraphics[scale=1]{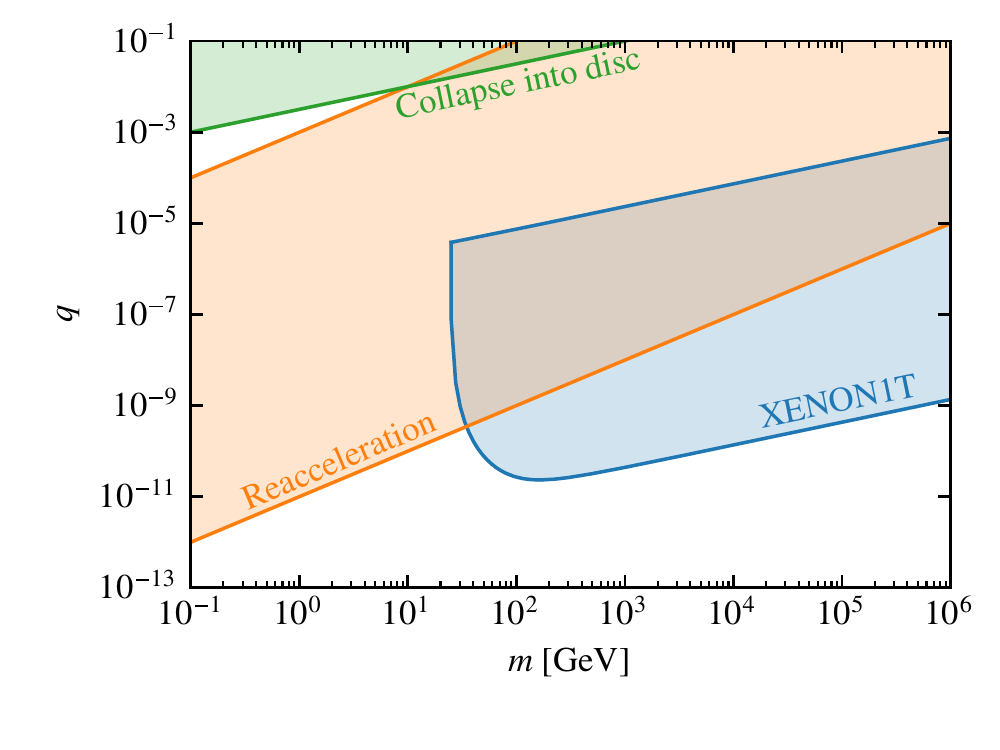}
  \caption{Exclusion plot of charged DM in the mass-charge plane for $f_{\rm cDM}=1$. 
  The orange area is excluded due to the unobserved drastic change in 
  the diffusion coefficient expected due to reacceleration of charged DM.
  The Xenon 1T 90\% exclusion (blue area) is obtained from Ref. 
  \cite{Belanger:2020gnr}, and the exclusion due to possible collapse into 
  the disc is obtained from Ref. \cite{Dunsky:2018mqs}.}
  \label{fig:upper_limit}
\end{figure}

In Fig.~\ref{fig:upper_limit}, we show the upper limit in the $(m, q)$
parameter space for $f_{\rm cDM}=1$ . The exclusion area for Xenon~1T is taken
from Ref.~\cite{Belanger:2020gnr}. Furthermore, charged DM with
$m/q^2\lesssim 10^5\unit{GeV}$  would have collapsed into the disc and is thus
excluded~\cite{Dunsky:2018mqs}. Additional constraints are discussed in, e.g.,
Refs.~\cite{Dunsky:2018mqs,McDermott:2010pa,Caputo:2019tms}.

Finally, we note that for $f_{\rm cDM}<1$ most existing exclusion limits become 
weaker. In the case of reacceleration, however, we found that the exclusion
is limited by the resonance condition and holds as long as
$f_{\rm cDM}^{-1}m/q \lesssim 10^{12}\unit{GeV}$. Allowing for a partial
transmission of wave-power to larger scales can further increase the excluded
region in $f_{\rm cDM}$. For $f_{\rm cDM}\lesssim 10^{-6}$, the acceleration
of charged DM will disturb its momentum distribution, and the 
methods to derive exclusion limits for charged cosmic rays worked out in
Ref.~\cite{Dunsky:2018mqs}  might be applied.

\section{Conclusions} 
\label{sec:conclusion}

In this work we have shown that the feedback due to second-order Fermi
acceleration
is an important effect that must be taken into account when analysing the
propagation of charged dark matter. 
The growth rate turns out to be dominant at mass to charge ratios
$m/q\lesssim 10^{15}\unit{GeV}$.
As such, the absorption of Alfv\'en waves will
stop the cascading of wave-power to smaller scales
around the Larmor radius of the charged particles, $k_{\rm res}\sim 1/R_L$.
This will in turn imply a significant and sudden
change in the diffusion coefficient for ordinary cosmic rays
at this scale.
This unobserved consequence
leads to the excluded region
$10^3\unit{GeV}\lesssim m/q \lesssim 10^{11}\unit{GeV}$.
Even more, this limit remains fixed as long as the charged dark matter abundance
$f_X=\Omega_{\rm cDM}/\Omega_{\rm DM}$ satisfies
$f_{\rm cDM}^{-1}m/q \lesssim 10^{12}\unit{GeV}$.


\begin{thebibliography}{10}

\bibitem{Ignatiev:1978xj}
A.~{\relax Yu}. Ignatiev, V.~A. Kuzmin and M.~E. Shaposhnikov, \emph{{Is the
  Electric Charge Conserved?}},
  \href{https://doi.org/10.1016/0370-2693(79)90048-0}{\emph{Phys. Lett.}
  {\bfseries 84B} (1979) 315}.

\bibitem{Okun:1983vw}
L.~B. Okun, M.~B. Voloshin and V.~I. Zakharov, \emph{{Electrical neutrality of
  atoms and Grand Unification models}},
  \href{https://doi.org/10.1016/0370-2693(84)91884-7}{\emph{Phys. Lett.}
  {\bfseries 138B} (1984) 115}.

\bibitem{Georgi:1983sy}
H.~Georgi, P.~H. Ginsparg and S.~Glashow, \emph{{Photon Oscillations and the
  Cosmic Background Radiation}},
  \href{https://doi.org/10.1038/306765a0}{\emph{Nature} {\bfseries 306} (1983)
  765}.

\bibitem{Holdom:1985ag}
B.~Holdom, \emph{{Two U(1)'s and Epsilon Charge Shifts}},
  \href{https://doi.org/10.1016/0370-2693(86)91377-8}{\emph{Phys. Lett.}
  {\bfseries 166B} (1986) 196}.

\bibitem{Dobroliubov:1989mr}
M.~I. Dobroliubov and A.~{\relax Yu}. Ignatiev, \emph{{Millicharged
  particles}}, \href{https://doi.org/10.1103/PhysRevLett.65.679}{\emph{Phys.
  Rev. Lett.} {\bfseries 65} (1990) 679}.

\bibitem{Fayet:1990wx}
P.~Fayet, \emph{{Extra U(1)'s and New Forces}},
  \href{https://doi.org/10.1016/0550-3213(90)90381-M}{\emph{Nucl. Phys. B}
  {\bfseries 347} (1990) 743}.

\bibitem{DeRujula:1989fe}
A.~De~Rujula, S.~Glashow and U.~Sarid, \emph{{Charged dark matter}},
  \href{https://doi.org/10.1016/0550-3213(90)90227-5}{\emph{Nucl. Phys. B}
  {\bfseries 333} (1990) 173}.

\bibitem{Bowman:2018yin}
J.~D. Bowman, A.~E.~E. Rogers, R.~A. Monsalve, T.~J. Mozdzen and N.~Mahesh,
  \emph{{An absorption profile centred at 78 megahertz in the sky-averaged
  spectrum}}, \href{https://doi.org/10.1038/nature25792}{\emph{Nature}
  {\bfseries 555} (2018) 67}
  [\href{https://arxiv.org/abs/1810.05912}{{\ttfamily 1810.05912}}].

\bibitem{Barkana:2018lgd}
R.~Barkana, \emph{{Possible interaction between baryons and dark-matter
  particles revealed by the first stars}},
  \href{https://doi.org/10.1038/nature25791}{\emph{Nature} {\bfseries 555}
  (2018) 71} [\href{https://arxiv.org/abs/1803.06698}{{\ttfamily 1803.06698}}].

\bibitem{Kovetz:2018zan}
E.~D. Kovetz, V.~Poulin, V.~Gluscevic, K.~K. Boddy, R.~Barkana and
  M.~Kamionkowski, \emph{{Tighter limits on dark matter explanations of the
  anomalous EDGES 21 cm signal}},
  \href{https://doi.org/10.1103/PhysRevD.98.103529}{\emph{Phys. Rev. D}
  {\bfseries 98} (2018) 103529}
  [\href{https://arxiv.org/abs/1807.11482}{{\ttfamily 1807.11482}}].

\bibitem{Munoz:2018pzp}
J.~B. Mu\~noz and A.~Loeb, \emph{{A small amount of mini-charged dark matter
  could cool the baryons in the early Universe}},
  \href{https://doi.org/10.1038/s41586-018-0151-x}{\emph{Nature} {\bfseries
  557} (2018) 684} [\href{https://arxiv.org/abs/1802.10094}{{\ttfamily
  1802.10094}}].

\bibitem{Hertzberg:2019bvt}
M.~P. Hertzberg and M.~Sandora, \emph{{Dark Matter and Naturalness}},
  \href{https://doi.org/10.1007/JHEP12(2019)037}{\emph{JHEP} {\bfseries 12}
  (2019) 037} [\href{https://arxiv.org/abs/1908.09841}{{\ttfamily
  1908.09841}}].

\bibitem{Liu:2019knx}
H.~Liu, N.~J. Outmezguine, D.~Redigolo and T.~Volansky, \emph{{Reviving
  Millicharged Dark Matter for 21-cm Cosmology}},
  \href{https://doi.org/10.1103/PhysRevD.100.123011}{\emph{Phys. Rev. D}
  {\bfseries 100} (2019) 123011}
  [\href{https://arxiv.org/abs/1908.06986}{{\ttfamily 1908.06986}}].

\bibitem{Kahlhoefer:2020rmg}
F.~Kahlhoefer and E.~Urdshals, \emph{{On dark atoms, massive dark photons and
  millicharged sub-components}},
  \href{https://arxiv.org/abs/2001.04492}{{\ttfamily 2001.04492}}.

\bibitem{Fabbrichesi:2020wbt}
M.~Fabbrichesi, E.~Gabrielli and G.~Lanfranchi, \emph{{The Dark Photon}},
  \href{https://arxiv.org/abs/2005.01515}{{\ttfamily 2005.01515}}.

\bibitem{Filippi:2020kii}
A.~Filippi and M.~De~Napoli, \emph{{Searching in the dark: the hunt for the
  dark photon}}, \href{https://doi.org/10.1016/j.revip.2020.100042}{\emph{Rev.
  Phys.} {\bfseries 5} (2020) 100042}
  [\href{https://arxiv.org/abs/2006.04640}{{\ttfamily 2006.04640}}].

\bibitem{Davidson:1991si}
S.~Davidson, B.~Campbell and D.~C. Bailey, \emph{{Limits on particles of small
  electric charge}},
  \href{https://doi.org/10.1103/PhysRevD.43.2314}{\emph{Phys. Rev.} {\bfseries
  D43} (1991) 2314}.

\bibitem{McDermott:2010pa}
S.~D. McDermott, H.-B. Yu and K.~M. Zurek, \emph{{Turning off the Lights: How
  Dark is Dark Matter?}},
  \href{https://doi.org/10.1103/PhysRevD.83.063509}{\emph{Phys. Rev. D}
  {\bfseries 83} (2011) 063509}
  [\href{https://arxiv.org/abs/1011.2907}{{\ttfamily 1011.2907}}].

\bibitem{Chuzhoy:2008zy}
L.~Chuzhoy and E.~W. Kolb, \emph{{Reopening the window on charged dark
  matter}}, \href{https://doi.org/10.1088/1475-7516/2009/07/014}{\emph{JCAP}
  {\bfseries 07} (2009) 014} [\href{https://arxiv.org/abs/0809.0436}{{\ttfamily
  0809.0436}}].

\bibitem{SanchezSalcedo:2010ev}
F.~Sanchez-Salcedo, E.~Martinez-Gomez and J.~Magana, \emph{{On the fraction of
  dark matter in charged massive particles (CHAMPs)}},
  \href{https://doi.org/10.1088/1475-7516/2010/02/031}{\emph{JCAP} {\bfseries
  02} (2010) 031} [\href{https://arxiv.org/abs/1002.3145}{{\ttfamily
  1002.3145}}].

\bibitem{Li:2020wyl}
J.-T. Li and T.~Lin, \emph{{Dynamics of millicharged dark matter in supernova
  remnants}},  \href{https://arxiv.org/abs/2002.04625}{{\ttfamily 2002.04625}}.

\bibitem{Dunsky:2018mqs}
D.~Dunsky, L.~J. Hall and K.~Harigaya, \emph{{CHAMP Cosmic Rays}},
  \href{https://doi.org/10.1088/1475-7516/2019/07/015}{\emph{JCAP} {\bfseries
  1907} (2019) 015} [\href{https://arxiv.org/abs/1812.11116}{{\ttfamily
  1812.11116}}].

\bibitem{1969ocr..book.....G}
V.~L. {Ginzburg} and S.~I. {Syrovatskii}, \emph{{The origin of cosmic rays}}.
  New York: Gordon and Breach, 1969.

\bibitem{Skilling:1975a}
J.~{Skilling}, \emph{{Cosmic ray streaming - I. Effect of Alfv{\'e}n waves on
  particles.}}, \href{https://doi.org/10.1093/mnras/172.3.557}{\emph{\mnras}
  {\bfseries 172} (1975) 557}.

\bibitem{Skilling:1975c}
J.~{Skilling}, \emph{{Cosmic ray streaming - III. Self-consistent solutions.}},
  \href{https://doi.org/10.1093/mnras/173.2.255}{\emph{\mnras} {\bfseries 173}
  (1975) 255}.

\bibitem{Melrose:1968}
D.~Melrose, \emph{The emission and absorption of waves by charged particles in
  magnetized plasmas}, {\emph{Astrophysics and Space Science} {\bfseries 2}
  (1968) 171}.

\bibitem{1979ApJ...230..373E}
J.~A. {Eilek}, \emph{{Particle reacceleration in radio galaxies.}},
  \href{https://doi.org/10.1086/157093}{\emph{\apj} {\bfseries 230} (1979)
  373}.

\bibitem{doi:10.1029/JA095iA09p14881}
Y.~Zhou and W.~H. Matthaeus, \emph{Models of inertial range spectra of
  interplanetary magnetohydrodynamic turbulence},
  \href{https://doi.org/10.1029/JA095iA09p14881}{\emph{Journal of Geophysical
  Research: Space Physics} {\bfseries 95} (1990) 14881}.

\bibitem{refId0}
{Ptuskin, V. S.} and {Zirakashvili, V. N.}, \emph{Limits on diffusive shock
  acceleration in supernova remnants in the presence of cosmic-ray streaming
  instability and wave dissipation},
  \href{https://doi.org/10.1051/0004-6361:20030323}{\emph{A\&A} {\bfseries 403}
  (2003) 1}.

\bibitem{Armstrong:1995zc}
J.~Armstrong, B.~Rickett and S.~Spangler, \emph{{Electron density power
  spectrum in the local interstellar medium}},
  \href{https://doi.org/10.1086/175515}{\emph{Astrophys. J.} {\bfseries 443}
  (1995) 209}.

\bibitem{Chepurnov:2009ym}
A.~Chepurnov and A.~Lazarian, \emph{{Extending Big Power Law in the Sky with
  Turbulence Spectra from WHAM data}},
  \href{https://doi.org/10.1088/0004-637X/710/1/853}{\emph{Astrophys. J.}
  {\bfseries 710} (2010) 853}
  [\href{https://arxiv.org/abs/0905.4413}{{\ttfamily 0905.4413}}].

\bibitem{Boulanger:2018zrk}
F.~Boulanger et~al., \emph{{IMAGINE: A comprehensive view of the interstellar
  medium, Galactic magnetic fields and cosmic rays}},
  \href{https://doi.org/10.1088/1475-7516/2018/08/049}{\emph{JCAP} {\bfseries
  08} (2018) 049} [\href{https://arxiv.org/abs/1805.02496}{{\ttfamily
  1805.02496}}].

\bibitem{Schekochihin:2007mw}
A.~Schekochihin, S.~C. Cowley, W.~Dorland, G.~Hammett, G.~Howes, E.~Quataert
  et~al., \emph{{Kinetic and fluid turbulent cascades in magnetized weakly
  collisional astrophysical plasmas}},
  \href{https://doi.org/10.1088/0067-0049/182/1/310}{\emph{Astrophys. J.
  Suppl.} {\bfseries 182} (2009) 310}
  [\href{https://arxiv.org/abs/0704.0044}{{\ttfamily 0704.0044}}].

\bibitem{Thornbury:2014bwa}
A.~Thornbury and L.~O. Drury, \emph{{Power requirements for cosmic ray
  propagation models involving re-acceleration and a comment on second order
  Fermi acceleration theory}},
  \href{https://doi.org/10.1093/mnras/stu1080}{\emph{Mon. Not. Roy. Astron.
  Soc.} {\bfseries 442} (2014) 3010}
  [\href{https://arxiv.org/abs/1404.2104}{{\ttfamily 1404.2104}}].

\bibitem{2011hea..book.....L}
M.~S. {Longair}, \emph{{High Energy Astrophysics}}. Cambridge: University
  Press, 2011.

\bibitem{2005ARA&A..43..337C}
D.~P. {Cox}, \emph{{The Three-Phase Interstellar Medium Revisited}},
  \href{https://doi.org/10.1146/annurev.astro.43.072103.150615}{\emph{\araa}
  {\bfseries 43} (2005) 337}.

\bibitem{Kachelriess:2019oqu}
M.~Kachelrie{\ss} and D.~Semikoz, \emph{{Cosmic Ray Models}},
  \href{https://doi.org/10.1016/j.ppnp.2019.07.002}{\emph{Prog. Part. Nucl.
  Phys.} {\bfseries 109} (2019) 103710}
  [\href{https://arxiv.org/abs/1904.08160}{{\ttfamily 1904.08160}}].

\bibitem{Belanger:2020gnr}
G.~Belanger, A.~Mjallal and A.~Pukhov, \emph{{Recasting direct detection limits
  within micrOMEGAs and implication for non-standard Dark Matter scenarios}},
  \href{https://arxiv.org/abs/2003.08621}{{\ttfamily 2003.08621}}.

\bibitem{Caputo:2019tms}
A.~Caputo, L.~Sberna, M.~Frias, D.~Blas, P.~Pani, L.~Shao et~al.,
  \emph{{Constraints on millicharged dark matter and axionlike particles from
  timing of radio waves}},
  \href{https://doi.org/10.1103/PhysRevD.100.063515}{\emph{Phys. Rev. D}
  {\bfseries 100} (2019) 063515}
  [\href{https://arxiv.org/abs/1902.02695}{{\ttfamily 1902.02695}}].

\end{thebibliography}

\providecommand{\href}[2]{#2}\begingroup\raggedright\endgroup

\end{document}